\documentclass[letterpaper, 10 pt, conference]{ieeeconf} 

\IEEEoverridecommandlockouts        
\usepackage{amssymb,amsmath,amstext,amsopn,amsfonts,algorithmic}
\usepackage[dvips]{epsfig}
\usepackage{graphics,ifpdf}
\usepackage{color}
\usepackage{mathrsfs}
\usepackage{stmaryrd}
\usepackage{graphicx}
\usepackage{epstopdf}
\usepackage{pstricks,pst-plot,pst-node,pst-circ,moredefs}
\usepackage{caption}
\usepackage{subcaption}
\usepackage{yfonts}
\usepackage[Symbolsmallscale]{upgreek}

\usepackage{cite}

\usepackage{empheq}
\newlength\dlf

\newcommand{\ddt}[1]{\frac{\mathrm{d} #1}{\mathrm{d}t}}

\newcommand{\SO}{\ensuremath{\mathrm{SO(3)}}}

\newcommand{\T}{^{\mbox{\small T}}}
\newcommand{\so}{\ensuremath{\mathfrak{so}(3)}}

\newcommand{\bR}{\ensuremath{\mathbb{R}}}

\newcommand{\diag}{\mbox{diag}}
\newcommand{\bbm}{\begin{bmatrix}}
\newcommand{\ebm}{\end{bmatrix}}
\newcommand{\bm}{\begin{matrix}}
\newcommand{\eim}{\end{matrix}}
\newcommand{\matl}{\left[ \begin{array}}
\newcommand{\matr}{\end{array} \right]}
\newcommand{\be}{\begin{equation}}
\newcommand{\ee}{\end{equation}}
\newcommand{\bea}{\begin{eqnarray}}
\newcommand{\eea}{\end{eqnarray}}
\newcommand{\beas}{\begin{eqnarray*}}
\newcommand{\eeas}{\end{eqnarray*}}
\newcommand{\nn}{\nonumber}

\newcommand{\cM}{\mathcal{M}}
\newcommand{\cU}{\mathcal{U}}

\newcommand{\lan}{\langle}
\newcommand{\ran}{\rangle}

\newcommand{\U}{U^{m}}
\newcommand{\mrm}{\mathrm}
\DeclareMathOperator{\expm}{{expm}}  			

\newtheorem{theorem}{Theorem}
\newtheorem{assumps*}{Assumptions}

\newtheorem{proposition}[theorem]{Proposition}

\title{\LARGE \bf
Mechatronics Architecture of Smartphone-Based Spacecraft ADCS using VSCMG Actuators
}

\author{Sasi Prabhakaran Viswanathan$^{1}$, Amit Sanyal$^{2}$ and Maziar Izadi$^{3}$
\thanks{$^{1}$Sasi and $^{3}$Maziar are PhD students in the Department Mechanical and Aerospace Engineering, NMSU, USA
        {\tt\small sashi} and {\tt\small mi@nmsu.edu}}%
\thanks{$^{2}$Amit Sanyal is with the Faculty of Mechanical and Aerospace Engineering, NMSU, USA
        {\tt\small asanyal@nmsu.edu}}%
}

\begin{document}

\maketitle
\thispagestyle{empty}
\pagestyle{empty}

\begin{abstract}
Hardware and software architecture of a novel spacecraft Attitude Determination and Control System 
(ADCS) based on smartphones using Variable Speed Control Moment Gyroscope (VSCMG) as actuator is proposed here. A spacecraft 
ground simulator testbed for Hardware-in-the-loop (HIL) attitude estimation and control with VSCMG 
is also described. The sensor breakouts with independent micro-controller units are used in the 
conventional ADCS units, which are replaced by a single integrated off-the-shelf 
smartphone. On-board sensing, data acquisition, data uplink/downlink, state estimation and 
real-time feedback control objectives can be performed using this novel spacecraft ADCS. 
The attitude control and attitude determination (estimation) schemes have appeared in prior 
publications, but are presented in brief here. Experimental results from running the attitude 
estimation (filtering) scheme with the ``onboard'' sensors of the smartphone in the HIL simulator 
are given. These results, obtained in the Spacecraft Guidance, Navigation and Control Laboratory 
at NMSU, demonstrate the excellent performance of this estimation scheme with the 
noisy raw data from the smartphone sensors.  

\end{abstract}

\section{INTRODUCTION}
Spacecraft Attitude Determination and Control System (ADCS) is considered to be the vital sub-system of all other spacecraft systems, as it enables the critical capability of attitude maneuvers such as pointing, re-orientation, stabilization etc. An ADCS is necessary for commercial satellites, as well as spacecraft with science and tactical missions like observation of solar system bodies, stars and galaxies, as well as features and objects on the Earth's surface. The ADCS system consists of the attitude determination and estimation component, and the attitude control system \cite{ADCS-wertz}. A novel mechatronics approach  is proposed for spacecraft ADCS that uses a Commercial Off-The-Shelf (COTS) smartphone with integrated sensors as on-board computer. This technique utilizes the smartphone\rq{}s inbuilt accelerometer, magnetometer and gyroscope as an Inertial Measurement Unit (IMU) 
for attitude determination and a Variable Speed Control Moment Gyroscope array for attitude 
control.  

The primary motivation for using an open source smartphone is to create a cost-effective, generic 
platform for spacecraft attitude determination and control, while not sacrificing on performance and 
fidelity. The PhoneSat mission of NASA's Ames Research Center demonstrated the application of 
COTS smartphones as the satellite's onboard computer with its sensors being 
used for attitude determination and its camera for Earth observation \cite{Phonesat1}. More recent 
satellites in the Phonesat series, Phonesat 2.4 and 2.5, successfully utilized smartphones for 
attitude control using reaction wheels \cite{phonesat2}. University of Surrey's Surrey Space Centre 
(SSC) and Surrey Satellite Technology (SSTL) developed STRaND-1, a 3U CubeSat containing a 
smartphone payload \cite{strand1,strand2}.

For a typical mechatronics system, a number of sensor breakouts including accelerometer, rate 
gyros, magnetometer, communication breakouts and GPS units are used as independent sensor 
breakouts. Signals from these breakouts are processed through an on-board microcontroller, 
which estimates the states and calculates the control signals based on the feedback control law, 
which are then sent to the actuators. 
 A more complex 
mechanical system like a spacecraft with internal actuators demands a microcontroller with more 
processing power, memory capacity and input/output ports, which can considerably increase the 
cost and mass budget of the system. Evolution of smartphones have been remarkable in the past 
few years with sophisticated processors, higher order system architecture, and high quality sensors integrated 
into a compact unit. These smartphones can be used as on-board CPU for performing HIL 
operations of more complex mechatronic systems like spacecraft with internal 
actuators. Some advantages of using smartphones, on-board are:
 \begin{enumerate}
  \item compact form factor with powerful CPU, GPU etc.,
 \item integrated sensors and data communication options,
 \item long lasting batteries: reduces total mass budget, 
 \item cheap price and open source software development kit.
 \end{enumerate}
 
The novel ADCS scheme developed here includes an attitude and angular velocity estimation 
scheme that is based on inertial directions and angular velocity of the spacecraft measured by 
sensors in the body-fixed frame of the smartphone. The state estimation scheme presented here 
has the following important properties: (1) the attitude is represented globally over the 
configuration space of rigid body attitude motion without using local coordinates or quaternions; 
(2) the scheme developed does not assume any statistics (Gaussian or otherwise) on the 
measurement noise; (3) no knowledge of the attitude dynamics model is assumed; and (4) the 
continuous and discrete-time filtering schemes presented here are obtained by applying the 
Lagrange-d'Alembert principle or its discretization~\cite{marswest} to a Lagrangian function that 
depends on state estimate errors obtained from vector measurements for attitude and angular 
velocity measurements. 

The spacecraft\rq{}s attitude is controlled by the variable speed control moment gyroscope 
(VSCMG) actuators. A general dynamics model of a spacecraft with VSCMGs is derived 
in \cite{msc13, dscc13}, and is adopted here. Since the configuration space of attitude motion of 
a spacecraft with internal actuators is a nonlinear manifold, the global dynamics of this system is 
treated using the formulation of geometric mechanics~\cite{marat, blbook}. This model is obtained 
using variational mechanics, and in the framework of geometric mechanics on the nonlinear state 
space of this system. This dynamics model relaxes  simplifying assumptions used in most of the 
existing literature on (VS)CMGs 
\cite{hughes1986spacecraft,wie:2008,schaub:09,tokar1979singular}. These simplifying 
assumptions, which are relaxed here, includes:
\begin{enumerate}
\item Center of Mass (CoM) of gimbal is perfectly aligned with rotor CoM along the gimbal axis ($\sigma=0$),
\item axisymmetric rotor,
\item both the gimbal \& the rotor-fixed coordinate frames are their corresponding principal 
axes frames,
\item rotor \& gimbal inertias are about their respective CoM,
\item gimbal frame structure has ``negligible" inertia, 
\item angular rate of the gimbal frame is ``negligible" compared to the 
rotor angular rate about its symmetry axis.
\end{enumerate}
Salient features of this VSCMG model are system adaptability (misalignment correction), and scalability.

 
 The ADCS mechatronics architecture comprises of a COTS smartphone running stock Android OS with 
self-contained ADCS routine, an ADK microcontroller board, ESC units along with the motor driven VSCMG arrays. The 
standalone mechatronics architecture performs the task of state sensing through embedded 
MEMS sensors, filtering, state estimation and implementation of feedback control law, to achieve 
the desired control objectives while maintaining active uplink/downlink with a remote ground control station.


%
%
%

\section{Attitude Dynamics of a Spacecraft with VSCMG actuators}
\label{sec:kindyn}
\subsection{Spacecraft Attitude Kinematics}
Let $R(t)$ denote the rotation matrix from the base body-fixed coordinate frame to an inertial coordinate frame. If $\Omega(t)$ is the total angular velocity of the base body with respect to the inertial frame and expressed in the base body frame, then the attitude kinematics of the base body is given by 
\be \dot R(t)= R(t) \Omega(t)^\times. \label{atkinbb} \ee
The spacecraft with a VSCMG has $5$ rotational degrees of freedom, which are described 
by the variables: gimbal angle ($\alpha$), rotor angle ($\theta$) and attitude $R$.

\subsection{Equations of motion}
\begin{figure}
     \centering
                \includegraphics[width=0.45\textwidth]{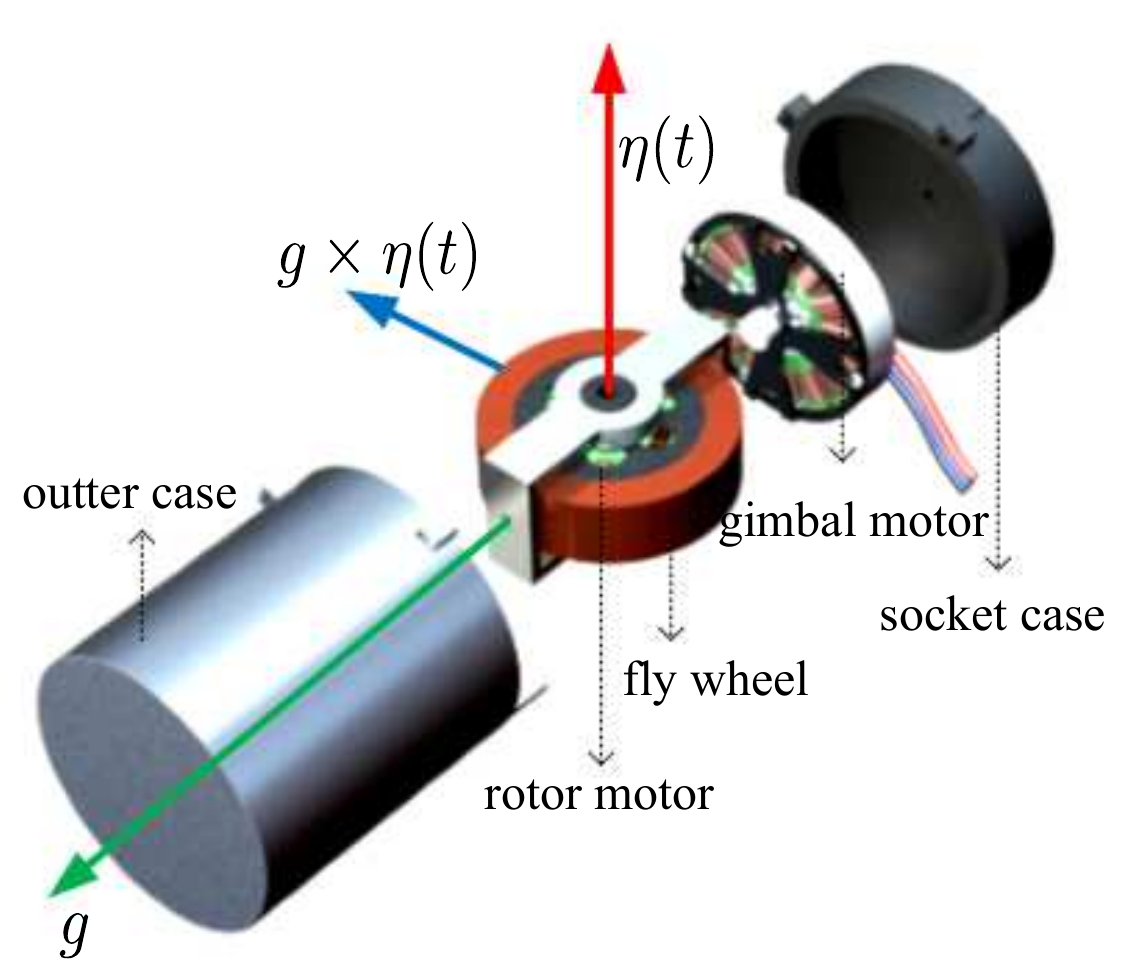}
        \caption {Exploded view of the VSCMG unit}\label{fig:vscmg_demo}
\end{figure}

The accurate dynamics model of a spacecraft with VSCMG derived in  \cite{msc13, dscc13}, 
appropriate for model-based controller design, is briefly presented in this section. The gimbal 
axis ($g$) of the VSCMG is orthogonal to the rotor's axis of rotation ($\eta(t)$), which passes through its CoM, if the rotor offset, $\sigma=0$. A schematic CAD rendering of a VSCMG assembly, is shown in Fig.~\ref{fig:vscmg_demo}. 
The VSCMG models accounts for rotor misalignments ($\sigma\neq0$) if any, which makes the 
system more adaptive and reliable. Total rotational kinetic energy of a spacecraft with a VSCMG is 
\be T=\frac 12 \chi(t)\T  \mathcal{J}(t) \chi(t) \label{kinET} \ee
where, 
\begin{align} 
\chi (t)&=\bbm \Omega(t) \\ \dot{\gamma}(t) \ebm \,\mbox{\& }\,
      \mathcal{J}(t) = \bbm J_T (t)+I_T(t) & B(t) \\ B(t)\T & J_{gr}(t), \ebm \\ 
B(t)&=\bbm    \big(J_c + I_c(t)\big)g &
R_r J_r R_r\T\eta(t)   \ebm \in \mathbb{R}^{3{\times} 2},  \\
\gamma (t)&=\bbm \alpha (t) \\ \theta (t) \ebm,\\
 J_{gr}(t)&= \bbm
       g\T J_c (t) g & g\T J_r\eta(t) \\
      \eta(t)\T J_r g & \eta(t)\T J_r \eta(t) \ebm \in \bR^{2\times 2} \label{Jgr}.
\end{align}
Define the following angular momentum quantities, which depend on the kinetic energy 
\eqref{kinET}:
\begin{align}
\mbox{ Total }\Rightarrow
\Pi  &= \frac{\partial T}{\partial\Omega}= (J_T+ I_T(t))
\Omega + B(t) \dot\gamma, \label{Pidef} \\
\mbox{ VSCMG }\Rightarrow
p &= \frac{\partial T}{\partial\dot\gamma}
= B(t) \T\Omega+ J_{gr}(t)  \dot\gamma. \label{pdef}
\end{align}

The equations of motion obtained from applying the Lagrange-d'Alembert principle can be 
written as follows:
\begin{empheq}[box=\fbox]{align}
\ddt{\Pi}=& \Pi\times\Omega + \cM_g(R(t)), \label{dyneq1} \\
\ddt{ p}=& \frac{\partial T}{\partial\gamma} + \tau. \label{dyneq2}
\end{empheq}
where, $ \cM_g(R(t))$ is the gravity gradient moment on the spacecraft, expressed in the base body frame and $\tau$ is the VSCMG control torque. 
The total angular momentum of the spacecraft can be expressed as the sum of the basebody momenta and the VSCMG momenta,
              \begin{align}
\Pi (t)=\overbrace{(J_T + I_T (t)) \Omega(t)}^\text{Spacecraft basebody momenta}+\underbrace{ B(t) \dot{\gamma} (t),}_\text{VSCMG momenta} \label{13}
\end{align}
\begin{align}
\mbox{where, }&J_T = J_b- (m_g+m_r)(\rho_g^\times)^2,  \nn \\
&I_T(t)= J_c(t) + I_c(t)- m_r\sigma \eta(t)^\times\rho_g^\times, \label{Pisimp}\\
&J_c(t)= R_g(t) J_g R_g(t)\T+R_r(t) J_r R_r(t)\T,\\ 
& I_c(t)= -m_r\sigma\big(\rho_g^\times\eta(t)^\times+\sigma(\eta(t)^\times)^2\big).  
\end{align}
Here, $J_b$ is the inertia of the spacecraft basebody; $J_g$ and $J_r$ are the corresponding 
gimbal and rotor inertias, respectively.  Total constant part of inertia in the base body frame is 
denoted by $J_T$ whereas $I_T(t)$ is the total time-varying part of inertia in the base body 
frame. $J_c(t)$ and $I_c(t)$ are the  offset-independent part of inertia and offset-dependent part 
of the VSCMG inertia in the spacecraft base body frame, respectively. Also, $R_{g}$ and $R_{r}$ 
denote the rotation matrix from the VSCMG gimbal-fixed and rotor-fixed coordinate frames to the 
spacecraft base body coordinate frame, respectively.

The dynamics of the base body is therefore affected by the rotation rates $\dot\alpha(t)$ and $\dot\theta(t)$ of the VSCMG gimbal and rotor respectively, but is independent of the rotation angle $\theta(t)$ of the rotor which is a cyclic variable for this system. Therefore, the conjugate momentum corresponding to $\theta(t)$, i.e. $p_\theta (t)$, is conserved if there is no torque input acting on the rotor axis. Note that the angular speed of the rotor $\dot\theta(t)$ is not constant in this case due to the dynamical coupling between the different degrees of freedom of this spacecraft system.
Therefore, to keep a constant rotor angular speed in the base spacecraft body frame as in constant speed SGCMG, a control torque has to be applied to the rotor.

%
%

\section{Spacecraft ADCS}
Spacecraft ADCS enables the spacecraft to realize its current states on \SO and can reorient the spacecraft to any desired final states. The raw  IMU measurements from the smartphone are fused/filtered through the estimation scheme given in Section \ref{Estimator}. Section \ref{controller} provides the control strategy for generic spacecraft attitude maneuvers.  

\subsection{Estimator}
\label{Estimator}
In \cite{Automatica}, an estimation of rigid body attitude and angular velocity without any knowledge of the attitude 
dynamics model, is presented using the Lagrange-d'Alembert principle from variational 
mechanics. This variational observer requires at least two body-fixed sensors to measure inertially known and constant direction vectors as well as sensors to read the angular velocity. A first order discretized estimation scheme for computer implementation is also presented using discrete variational mechanics. Here, a second order symmetric Lie group variational integrator is introduced.

In order to determine three-dimensional rigid body attitude instantaneously, three known inertial vectors are needed. This could be satisfied with just two vector measurements. In this case, the cross product 
of the two measured vectors is considered as a third measurement for applying the attitude 
estimation scheme. Let these vectors be denoted as $u_1^m$ and $u_2^m$, in the body-fixed frame. Denote the corresponding known inertial vectors as seen from the rigid body as $e_1$ and $e_2$, and 
let the true vectors in the body frame be denoted $u_i=R \T e_i$ for $i=1,2$, where $R$ is the rotation 
matrix from the body frame to the inertial frame. This rotation matrix provides a coordinate-free,  
global and unique description of the attitude of the rigid body. Define the matrix composed 
of all three measured vectors expressed in the body-fixed frame as column vectors, $U^m= [u_1^m\ u_2^m\ u_1^m\times u_2^m]$ 
and the corresponding matrix of all these vectors expressed in the inertial frame as
$E= [e_1\ e_2\ e_1\times e_2]$.
Note that the matrix of the actual body vectors $u_i$ corresponding to the inertial vectors 
$e_i$, is given by $U= R \T E= [u_1\ u_2\ u_1\times u_2]$.

\begin{proposition}\label{dis2ndfilter}
A discrete-time filter that gives a second order numerical integrator for the filter in \cite{Automatica} is given as follows:
\begin{align}
\begin{split}
&\hat R_{i+1}=\hat{R_i}\exp\big(h(\Omega_{i+\frac12}^m-\omega_{i+\frac12})^\times\big),\vspace*{4mm}\\
&m\omega_{i+1}=\exp(-\frac{h}{2} \hat\Omega_{i+1}^\times)\Big\{(m I_{3\times3}-\frac{h}{2}D)
\omega_{i+\frac12}\\
&~~~~~~~~~~~~+\frac{h}{2}\Phi'\big(\cU^0(\hat R_{i+1},\U_{i+1})\big)S_{L_{i+1}}(\hat R_{i+1})\Big\},
\vspace*{4mm}\\
&\hat\Omega_i=\Omega_i^m-\omega_i,\;\;\Omega_{i+\frac12}^m=\frac12(\Omega_i^m+\Omega_{i+1}^m),
\end{split}
\label{dfilteqns2nd}
\end{align}
\begin{align}
\mbox{where, }\omega_{i+\frac12}=& \big(mI_{3\times3}+\frac{h}{2}D\big)^{-1}\bigg\{ \exp\big(-\frac{h}{2}
\hat\Omega_i^\times\big) m\omega_i\nn\\
&+\frac{h}{2} \Phi'\big(\cU^0(\hat R_i,\U_i)\big)S_{L_i}(\hat R_i) \bigg\} \label {iAV}
\end{align}
is a discrete-time approximation to the angular velocity at time $t_{i+\frac12}:=t_i+\frac{h}{2}$,
where $h$ is the time step, $m$ is a positive scalar and $D$ is a positive definite matrix, $S_{L_i}(\hat R_i)=\mrm{vex}(L_i\T\hat R_i-\hat R_i\T L_i)\in\bR^3$ and $\mrm{vex}(\cdot): \so\to\bR^3$ is the inverse of the $(\cdot)^\times$ map, $L_i=E_i W_i(\U_i)\T\in\mathbb{R}^{3\times3}$ and $(\hat R_0,\hat\Omega_0)\in\SO\times\bR^3$ are initial estimated states. Further, $W$ and $\Phi(x)$ are chosen to satisfy the conditions in \cite{Automatica}, and
\begin{align}
\cU^0 (\hat R_i,\U_i)=\frac{1}{2}\lan E_i-\hat R_i \U_i,(E_i-\hat R_i \U_i)W_i\ran,
\label{dattindex}
\end{align}
denotes the artificial potential term (Wahba's cost function) which is generalized by applying the function $\Phi$ on it.
\end{proposition}

\subsection{Attitude Control using VSCMG array}
\label{controller}
From \eqref{13} one can express, 
\be \Pi=\Pi_b+u, \mbox{ where } \Pi_b=\Lambda(t)\Omega\, \mbox{ and }\, 
u= \mathcal{B}\, \dot{\Gamma}. \label{sc-cmg}\ee 
	
	The idea behind controlling the spacecraft attitude dynamics with the VSCMG is to ensure 
that the angular velocity of the spacecraft, which depends only on the term $\Pi_b$ in 
equation \eqref{sc-cmg}, is controlled using the internal momentum $u$, which depends 
on the VSCMG rotation rates $\dot\Gamma$. For simplicity, we consider the case where 
there is no gravity gradient moment, i.e., $M_g(R(t))\equiv 0$. Therefore, re-expressing the 
base-body attitude dynamics \eqref{sc-cmg} as:
\begin{align}
\dot{\Pi}_b&=\Pi_b \times \Omega+ (u \times \Omega-\dot u) 
	              =\Pi_b \times \Omega+ \tau_{cp},
\label{contau}
\end{align}
where $\tau_{cp}= u \times \Omega-\dot u$ is the control torque generated by the ``internal"
momentum $u$ from the VSCMGs. Note that from \eqref{contau} and the attitude kinematics 
\eqref{atkinbb}, the total momentum in an inertial frame, $\Pi_I= R\Pi= R(\Pi_b+u)$ 
is conserved. There are several control problems of interest for a spacecraft
with n VSCMGs. For most applications, the common spacecraft attitude maneuvers are slewing to rest, pointing , and attitude tracking maneuver. The results and discussion for a slew to rest attitude control scheme (de-tumbling maneuver) and a pointing maneuver for a spacecraft with $n$ VSCMGs are presented in references \cite{msc13, dscc13} and \cite{AAS14} respectively.

\section{Mechatronics architecture}
\subsection{Hardware}
\begin{figure}[h!]
 \centerline{\includegraphics[width=0.4\textwidth]{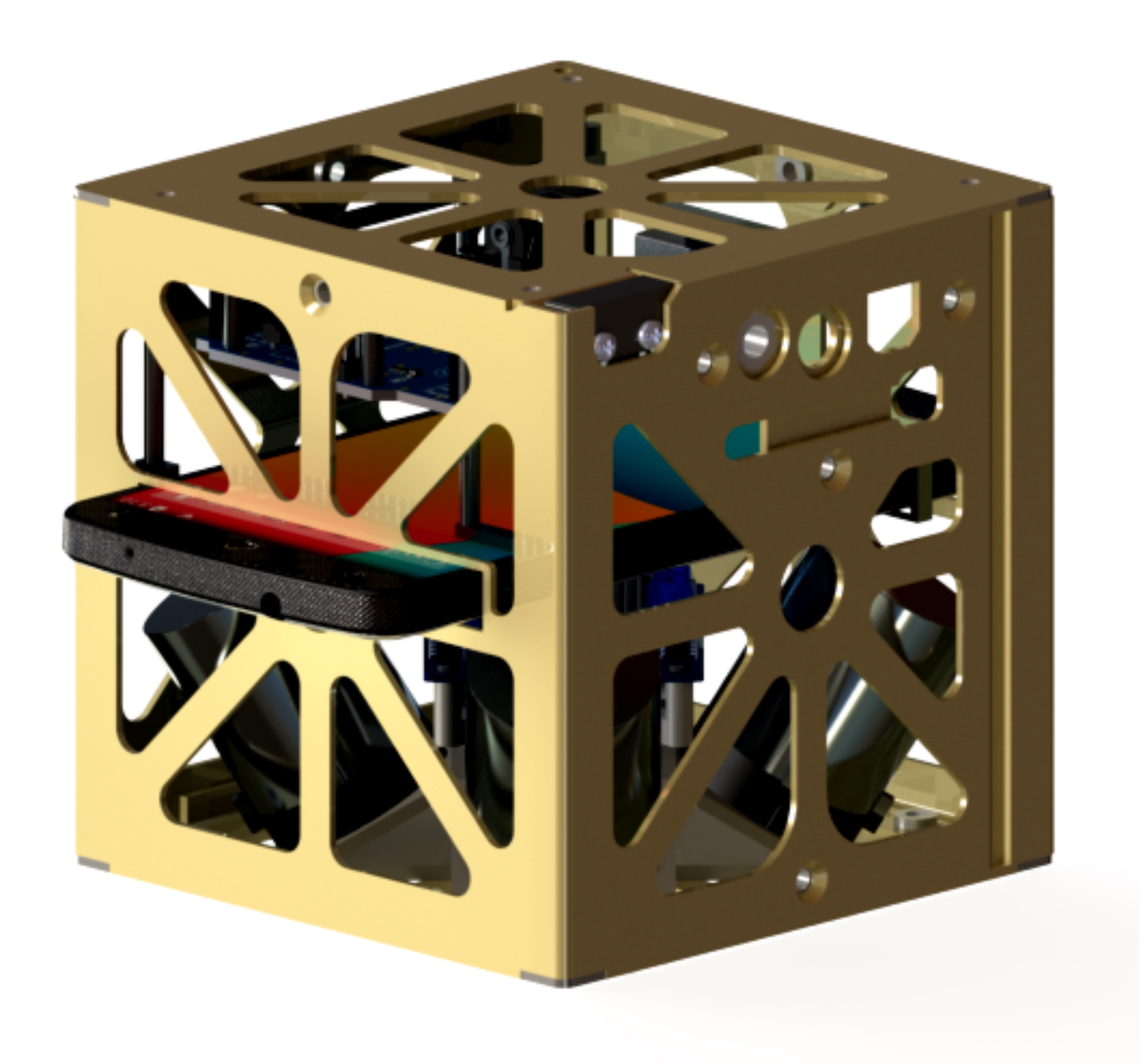}}
\caption{Spacecraft bus with smartphone and ADCS}
\label{scbus}
\end{figure}

For demonstrating the technological feasibility of the smartphone based ADCS using VSCMG, a 1U cubesat with miniature VSCMG actuator array as shown in Fig.~\ref{scbus} is considered, nevertheless the VSCMG system is scalable to any operational range from cubesat to space station. Each hermetically sealed VSCMG unit houses two brushless DC motor individually sized, controlled to drive the gimbal and rotor. An exploded of VSCMG showing its internal components is depicted in Fig.~\ref{fig:vscmg_demo}. The VSCMG array is formed by arranging $4$ individual units in pyramid configuration, where its skew angle is given by $\beta$. The main advantage of using VSCMG as attitude actuator is to exploit  its output torque to the input power ratio properties.       

Modern smartphones are integrated with the following hardware sensors:
\begin{enumerate}
\item Accelerometer
\item Gyroscope
\item Magnetometer
\item GPS, 5) Barometer, 6)  Light and proximity sensor etc.
\end{enumerate}
Out of these hardware sensors, ($1$), ($2$)  and ($4$) are fused and used as an Inertial Measurement Unit (IMU) for attitude determination. Sensors ($3$) and ($5$) can be used for altitude measurements in UAVs. The estimation and controller routine embedded in the on-board smartphone computes the required output signal and commands the VSCMG array to generate the corresponding actuator torque via Electronic Speed Control (ESC). The control signal from the on-board spamrtphone is sent to an microcontroller enabled with Accessory Development Kit (ADK) compatibility (eg. Google ADK, Arduino MEGA ADK, IOIO, Microbridge etc.), where corresponding Pulse Position Modulation (PPM) signals are generated and sent to ESC to control each VSCMG parameters, i.e. gimbal angle ($\alpha$), gimbal rate ($\dot{\alpha}(t)$), rotor angle ($\theta$) and rotor rate ($\dot{\theta(t)}$). The ESC units along with an Universal Battery Elimination Circuit (UBEC) are independently calibrated and programmed to drive each gimbal and rotor motors of the VSCMG array. The microcontroller interfaces the smarphone with other peripherals including  ADCS array and payloads; in addition to the system watchdog service.    
  
\begin{figure}[h]
 \centerline{
\includegraphics[width=0.5\textwidth]{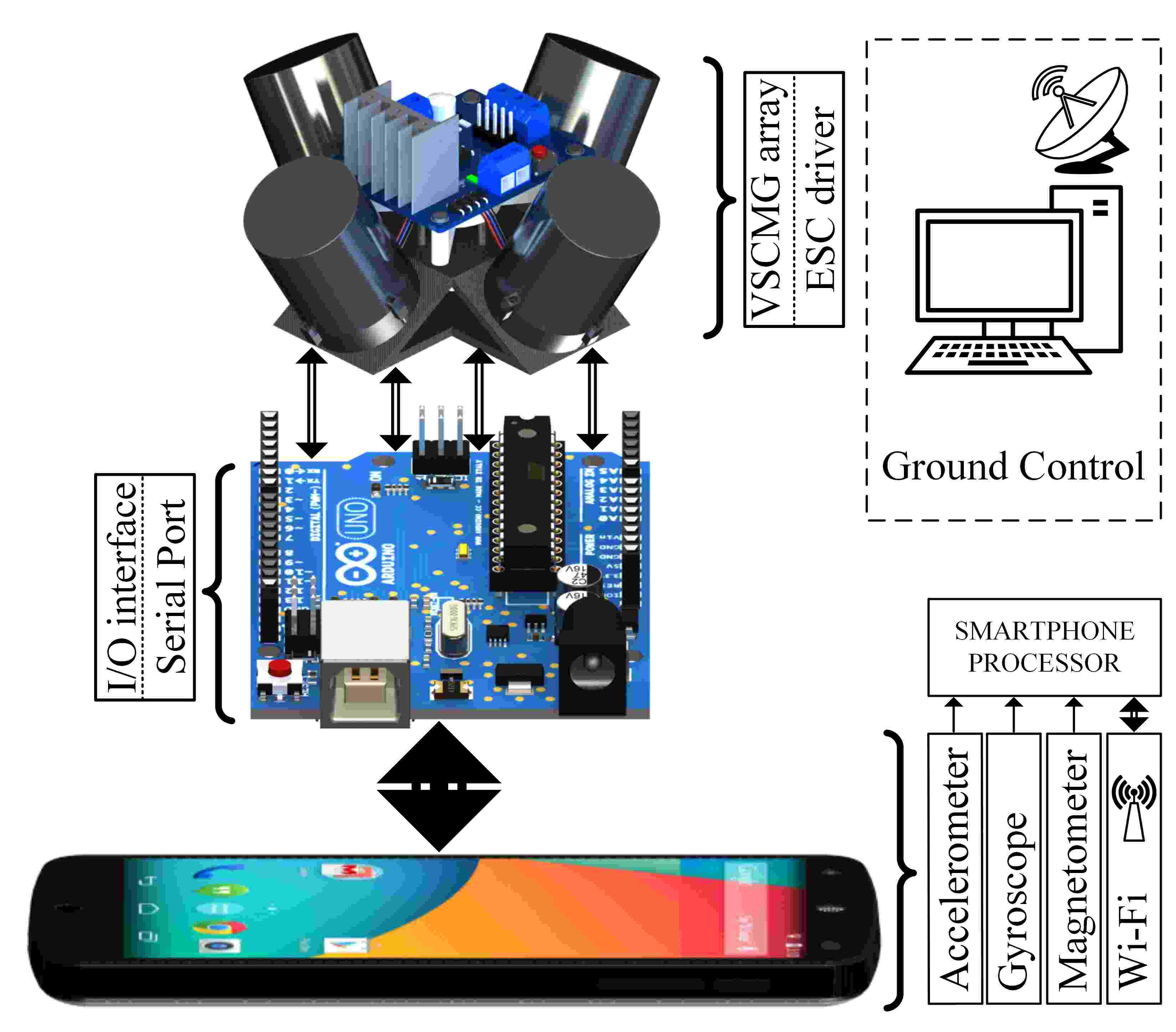}}
\caption{Hardware architecture of smartphone based ADCS}
\label{hard}
\end{figure}

\subsection{Software}
\subsubsection{Smartphone}
\begin{figure}[h]
 \centerline{
\includegraphics[width=0.5\textwidth]{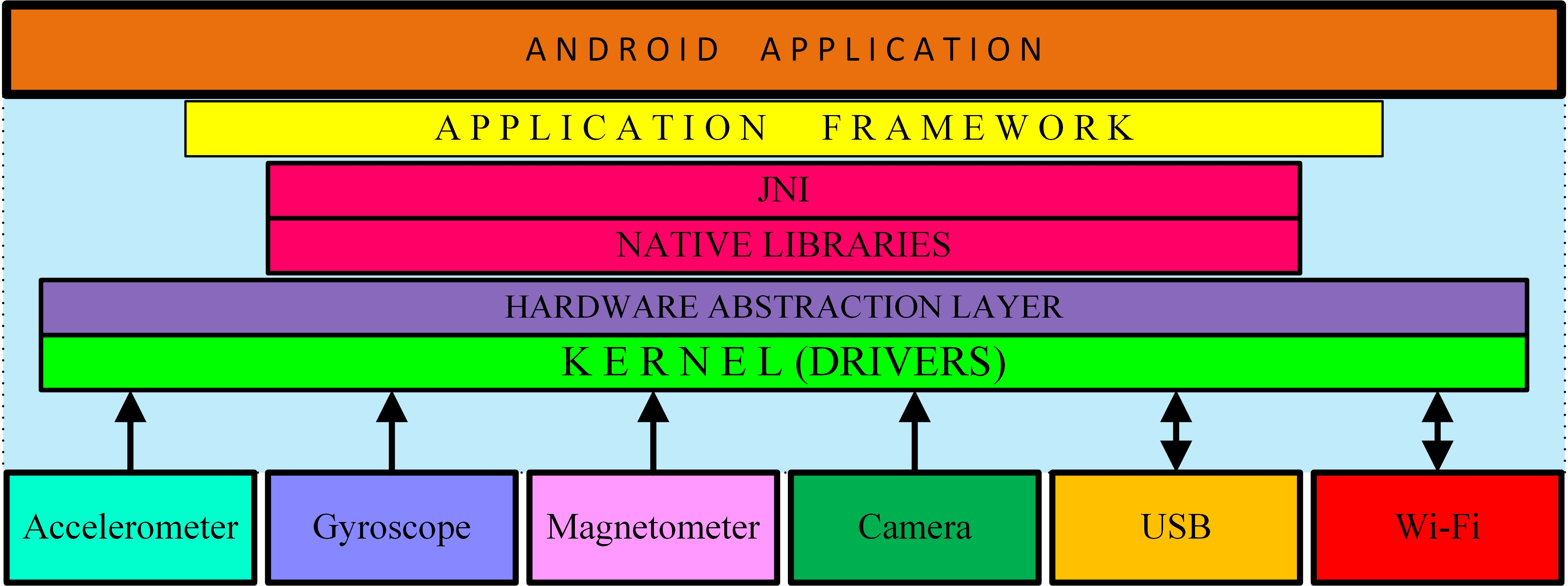}}
\caption{low level architecture}
\label{archi}
\end{figure}

Android is a Linux-based, open source mobile operating system that is very popular among 
software developers because of its powerful capabilities and open architecture. ADCS processes 
like sensor data acquisition, pre-filtering, sensor data fusion, estimation and control schemes 
are executed on-board the smarthphone\rq{}s processor. These functions are run as embedded 
native machine code compiled from C and/or C++ source files in the Android Application Packages 
($*.$apk) using the Android Native Development Kit (NDK) toolset \cite{ratabouil2012android}. 
Android NDK-based spacecraft ADCS programming is especially useful because it enables 
carrying out CPU-intensive workloads in real-time. A low level architecture shown in Fig.~\ref{archi}, 
is exploited for implementing the data acquisition from the accelerometer, gyroscope and 
magnetometer sensors. This is then used by the estimation scheme for attitude and angular 
velocity estimation through the Android Java application that interacts with native code using the 
Java Native Interface (JNI). External sensor breakouts and payloads are coupled to the 
application framework through the stable Linux subsystem I2C by writing the drivers in kernel 
and linking to the Hardware Abstraction Layer (HAL).
   
\subsubsection{Microcontroller}
The Accessory Development Kit (ADK) is a microcontroller development board that adheres to 
the simple Open Accessory Standard Protocol created by Google as a reference 
implementation \cite{bohmer2012beginning}. The ADK 2012 is based on the Arduino open 
source electronics prototyping platform, with some hardware and software extensions that allow it 
to communicate with Android devices. The ADK microcontroller has embedded within it the system 
program written in the C$++$ IDE, in parallel to the 
Android side programming. In addition, a sophisticated power management scheme can be 
implemented in the microcontroller, as all the system components including the smartphone 
and VSCMG actuators are powered by external battery packs via the UBEC and microcontroller. 

\section{HIL simulation of spacecraft attitude control}
\label{HIL}
\subsection{Testbed Hardware}
The spacecraft  attitude simulator testbed uses hemispherical air--bearing  to perform unrestricted yaw motion and restricted pitch and roll motion. The spacecraft bus is mounted on the air--bearing assembly and exhibits the same rotational degree of freedom (DoF) as that of the air--bearing. The spacecraft bus is actuated by a cluster of three VSCMG  arranged in tetrahedron configuration. Each VSCMG is independently driven by a high torque stepper motor to drive the gimbal and a variable speed servo motor to drive the rotor.  
\begin{figure}[h!]
 \centerline{
\includegraphics[width=0.4\textwidth]{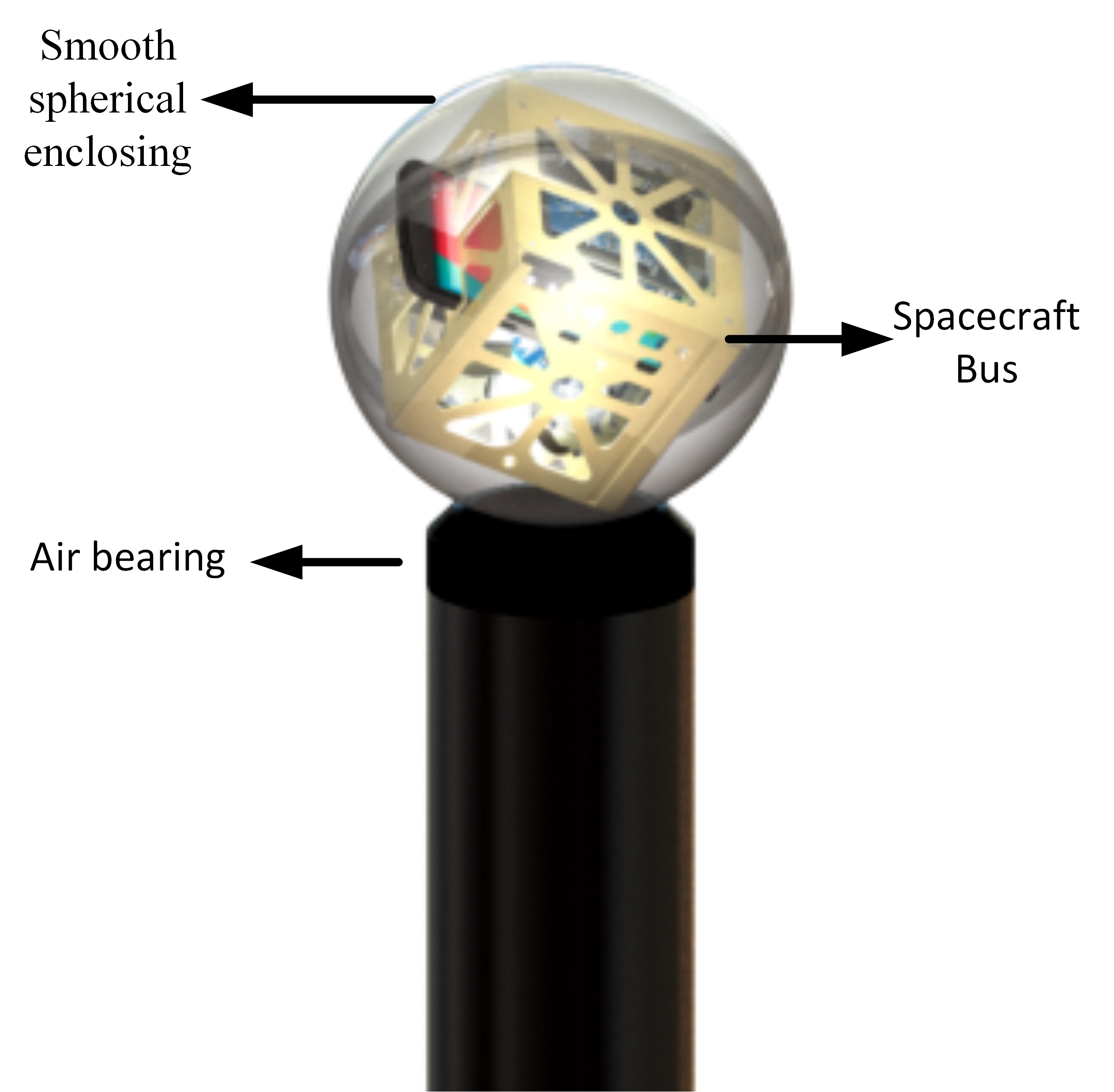}}
\caption{HIL testbed}
\end{figure}

\subsection{Experiments}
The estimation scheme given in Section \ref{Estimator} is implemented off-board on a remote PC using the sensor measurement acquired and transmitted by the on-board smartphone. The coordinates used as the inertial frame is ENU, which is a right-handed Cartesian frame formed by local east, north and up. The coordinates fixed to the COM of the cellphone with right direction of the screen as $x$, up direction as $y$ and the direction out of screen as $z$ is considered to be the body fixed frame. As mentioned in \ref{Estimator}, at least two inertially known and constant directions are required in order to estimate the rigid body attitude. Given the sensors installed on the smartphone, one could use the accelerometer to measure the gravity direction and the magnetometer to find the geomagnetic field direction. The cross product of these two vectors is considered as the third vector. In order to find these directions, one could normalize the data from accelerometer and magnetometer in the case that the cellphone is aligned with the true geographical directions and the body fixed frame coincides with the ENU frame. Note that the direction read by the accelerometer shows the up direction, since an upward acceleration equal to $g$ is applied to the phone in order to cancel the Earth's gravity and keep the phone still. Therefore, the matrix of three inertially constant directions is expressed in ENU frame as
$$E=[e_1\ e_2\ e_1\times e_2]=
\begin{bmatrix}
0 & 0.0772 & -0.9921\\
0 & 0.6117 & 0.1251\\
1 & -0.7873 & 0
\end{bmatrix}.
$$
The three axis gyroscope also gives the angular velocity measurements. These three sensors  produce measurement data at different frequencies. The filter's time step is selected according to the fastest sensor, which is accelerometer here. At those time instants where some of the sensors readings are not available because of the difference in sampling frequencies, the last read value 
from that sensor is used.

$\Phi(\cdot)$ could be any $C^2$ function with the properties described in Section 2 of \cite{Automatica}, but is 
selected to be $\Phi(x)=x$ here. Further, $W$ is selected based on the value of $E$, such that it satisfies the conditions in \cite{Automatica} as below:
$$W=
\begin{bmatrix}
3.19 & 1.51 & 0\\
1.51 & 3.19 & 0\\
0 & 0 & 2
\end{bmatrix}.
$$
The inertia scalar gain is $m=0.5$ and the dissipation matrix is selected as the following positive definite matrix:
\[D=\diag\big([12\ 13\ 14]\T\big).\]

Sensors outputs usually contain considerable levels of noise that may harm the behavior of the nonlinear filter. A Butterworth pre-filter is implemented in order to reduce these high-frequency noises. Note that the true quantities would not contain high-frequency signals, since they are related to a rigid body motion. A symmetric discrete-time filter for the first order Butterworth pre-filter is implemented for filtering the measurement data as follows:
\begin{align}
(2+ h)\bar{x}_{k+1}= (2-h)\bar{x}_k + h (x^m_k + x^m_{k+1}),
\end{align}
where $h$ is the time stepsize, $\bar{x}$ and $x^m$ are the filtered and measured quantities, respectively, and the subscript $k$ denotes the $k$th time stamp.
The initial estimated states have the following initial estimation errors:
\begin{align}
Q_0=\expm&_{\SO}\bigg(\Big(2.2\times[0.63\;\;\; 0.62\ -0.48]\T\Big)^\times\bigg), \nn\\
\mbox{and } \omega_0&=[0.001\ 0.002\ -0.003]\T\mbox{ rad/s}.
\end{align}
In order to integrate the implicit set of equations in \eqref{dfilteqns2nd} numerically, the first equation is solved at each sampling step, then the result for $\hat R_{i+1}$ is substituted in the second one. Using the Newton-Raphson method, the resulting equation is solved with respect to $\omega_{i+1}$ iteratively. The root of this nonlinear equation with a specific accuracy along with the $\hat R_{i+1}$ is used for the next sampling time instant. This process is repeated to the end of the simulation time.
Using the aforementioned quantities and the integration method, the simulation is carried out.

\subsection{Results}
\begin{figure}
\begin{center}
\includegraphics[width=0.5\textwidth]{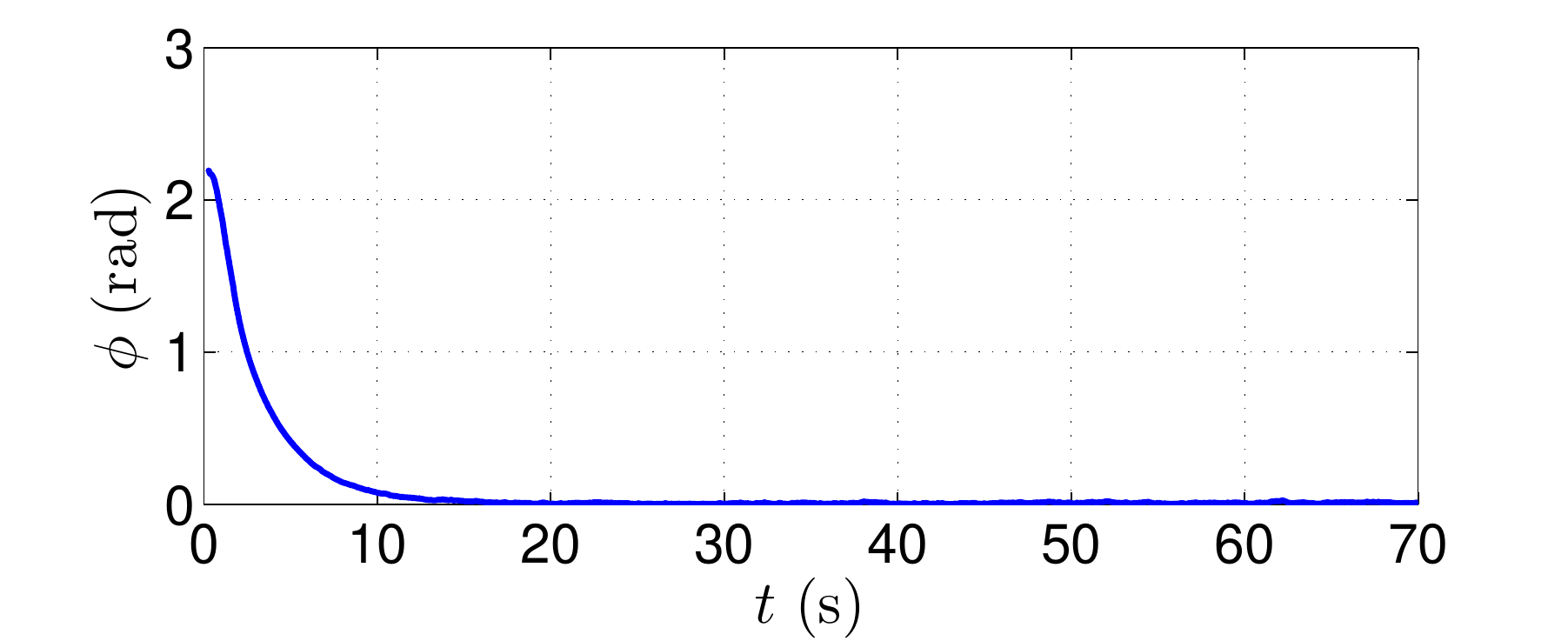}    
\caption{Principle Angle of the Attitude Estimation Error}  
\label{fig1}                                 
\end{center}                                 
\end{figure}

\begin{figure}
\begin{center}
\includegraphics[width=0.5\textwidth]{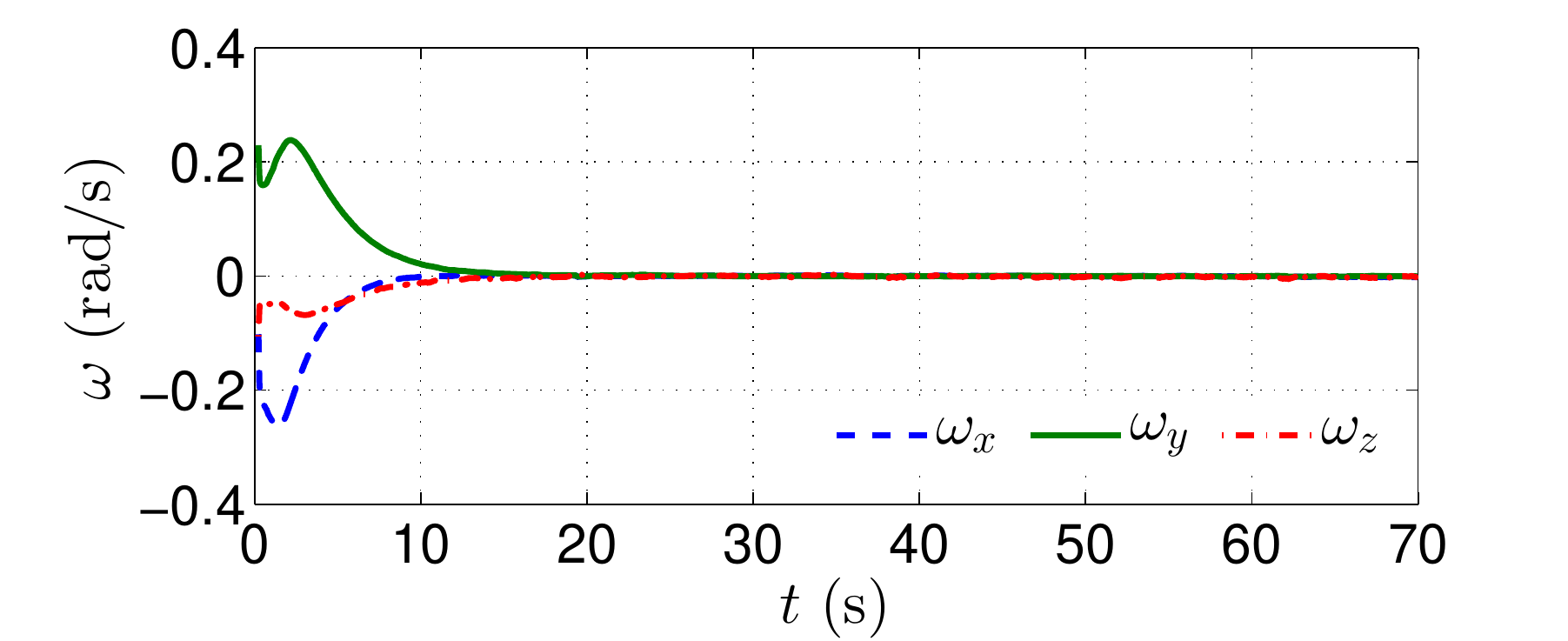}    
\caption{Angular Velocity Estimation Error}  
\label{fig2}                                 
\end{center}                                 
\end{figure}

Experimental results for the attitude estimation scheme, obtained from the experimental 
setup described in the previous subsection, are presented here. These experiments were 
carried out on the HIL simulator testbed in the Spacecraft Guidance, Navigation and Control 
laboratory at NMSU's MAE department. Experimental results from the control are not available 
yet since the VSCMG actuators have not been finalized and mounted on this testbed. 
The principal angle corresponding to the rigid body's attitude estimation error is depicted in Figure
\ref{fig1}. Estimation errors in the rigid body's angular velocity components are shown in 
Figure \ref{fig2}. All the estimation errors are seen to converge to a neighborhood 
of $(Q,\omega)=(I,0)$, where the size of this neighborhood depends on the characteristics of the 
measurement noise.

\section{Conclusion}

This paper presents a novel overall (hardware and software) architecture of a spacecraft 
attitude determination and control subsystem (ADCS), using a smartphone as the onboard 
computer. This architecture is being implemented using a HIL ground simulator for three-axis 
attitude motion simulation in the Spacecraft Guidance, Navigation and Control laboratory at 
NMSU. Theoretical and numerical results for the attitude control and attitude estimation schemes 
that are part of this architecture, have appeared in recent publications. Both the attitude control 
scheme and the attitude estimation scheme provide almost global asymptotic stability, and are 
robust to measurement noise and bounded disturbance inputs acting on the spacecraft. 
Experimental verification of the attitude estimation algorithm is presented here, and the 
experimental results show excellent agreement with the theoretical and numerical results 
on this algorithm that have appeared in recent publications.

\end{document}